\documentclass[a4paper,twocolumn,10pt]{article}
  \usepackage[left=1.8cm, right=1.8cm]{geometry}
	\usepackage[numbers]{natbib}
  \usepackage{tabularx}
	\usepackage{amssymb}  
	\usepackage{amsmath}  
	\usepackage[english]{babel}   
	\usepackage[all,cmtip]{xy} 
	\DeclareMathOperator{\h}{\mathcal{H}} 
	\DeclareMathOperator{\A}{\mathcal{A}} 
	\DeclareMathOperator{\een}{1\hspace{-0.25em}\mathrm{l}} 
	\DeclareMathOperator{\dee}{d} 
	\DeclareMathOperator{\Trace}{Tr} 
	\DeclareMathOperator{\pee}{I\hspace{-0.2em}\mathrm{P}} 
	\DeclareMathOperator{\alsdan}{\textbf{=}\hspace{-0.8em}>} 
	\DeclareMathAlphabet{\mathpzc}{OT1}{pzc}{m}{it} 
	\newtheorem{theorem}{Theorem}[section]
	
	\usepackage[T1]{fontenc} 
\begin{document}

\title{The Problem of Contextuality and the Impossibility of Experimental Metaphysics Thereof} 
\author{Ronnie Hermens}

\setlength{\unitlength}{\textwidth}
\twocolumn[
  \begin{@twocolumnfalse}
    \maketitle
    \begin{abstract}
Recently a new impulse has been given to the experimental investigation of contextuality.
In this paper we show that for a widely used definition of contextuality there can be no decisive experiment on the existence of contextuality. To this end, we give a clear presentation of the hidden variable models due to Meyer, Kent and Clifton (MKC), which would supposedly nullify the Kochen-Specker Theorem. Although we disagree with this last statement, the models play a significant role in the discussion on the meaning of contextuality. In fact, we introduce a specific MKC-model of which we show that it is non-contextual and completely in agreement with quantum mechanical predictions. We also investigate the possibility of other definitions of non-contextuality --with an emphasis on operational definitions-- and argue that any useful definition relies on the specification of a theoretical framework. It is therefore concluded that no experimental test can yield any conclusions about contextuality on a metaphysical level. 
\end{abstract}
 \end{@twocolumnfalse}
  ]



\tableofcontents 
 
\section{Introduction}
It is the general consensus that the Kochen-Specker Theorem excludes the existence of non-contextual hidden variable theories. Naturally this leads to the questions of what is really meant by the notion of contextuality, what the implications are for the future of hidden variable theories, and if experimental proof of this statement is possible. Over the past ten years the discussion on these questions has had an enormous boost but, surprisingly, with mutually opposing outcomes results. On the one hand Meyer, Kent, Clifton and Barrett \cite{Barrett-Kent04}, \cite{CliftonKent99}, \cite{Kent99}, \cite{Meyer99} have advocated the view that non-contextual hidden variable models can be constructed that reproduce the quantum mechanical predictions. On the other hand there have been experiments that claim to prove that nature is contextual, in accordance with quantum mechanical predictions \cite{Bartosik09}, \cite{Guhne10}, \cite{Huang03}, \cite{Kirchmair09}, \cite{Liu09}, \cite{Michler00}, \cite{Moussa10}, \cite{Pan10}, \cite{Spekkens09}. Thus far, no serious attempt has been made to clarify the paradox at hand. That is, until now.

In section \ref{MKC-sectie} of this paper we give a clear presentation of the non-contextual hidden variable models proposed by Meyer, Kent and Clifton. These models turn out to be very flexible and we introduce one that is completely empirically equivalent with quantum mechanics. Our hope is that we will remove a lot of the misconceptions about these models, which have led to objections against these models, cf. \cite{Barrett-Kent04}, \cite{Hermens10}. 
In this paper these models will play a leading role in the investigation of several notions of non-contextuality and the possibility of experimental tests thereof. 

In section \ref{operational-sectie} we study the operational definitions of non-contextuality as proposed in \cite{Spekkens05} and how the MKC-models behave with respect to these definitions. We will argue that whether or not the MKC-models are contextual with respect to these definitions relies on the precise interpretations of these definitions. This demonstrates that operational non-contextuality is rather ambiguously defined. 

In section \ref{experimental-sectie} we will shed a light on what is actually shown in several proposed and conducted tests of contextuality. It will be shown that the tests of common notions of contextuality depend on auxiliary assumptions that aren't satisfied by the MKC-models. In section \ref{Spekkenstest} it is shown that the experimental tests of operational non-contextuality rely on an interpretation of the definition of non-contextuality for which even classical physics turns out to be contextual. This is taken as a strengthening of our opinion that these definitions are ambiguous.  
Finally, we take a more general viewpoint on the idea of experimental tests of contextuality and argue that there can be no notion of contextuality that is both metaphysically interesting and experimentally accessible.

\section{The MKC-models as a non-contextual hidden variable theory}\label{MKC-sectie}
\subsection{The Kochen-Specker Theorem}

The models of Meyer, Kent and Clifton (MKC, \cite{CliftonKent99}, \cite{Kent99}, \cite{Meyer99}) are constructed around the Kochen-Specker Theorem \cite{KS67}. Therefore, a clear description of this theorem is mandatory for a good understanding of the models. We present it here as a purely mathematical result so as to keep the philosophical and the mathematical discussions as much separated as possible.

\begin{theorem}[Kochen \& Specker 1967]\label{KS-Theorem}
Let $\h$ be a Hilbert space and let $B_{sa}(\h)$ 
denote the set of self-adjoint operators on this space with domain dense in $\h$. Then, if $\dim(\h)>2$, there exists no function $\lambda:B_{sa}(\h)\to\mathbb{R}$ such that
\begin{enumerate}
	\item for every $A\in B_{sa}(\h)$: $\lambda(A)$ lies in the spectrum of $A$; 
	\begin{equation}\sigma(A)=\{x\in\mathbb{R}\:;\:A-x\een\text{ not invertible}\},\end{equation}
	\item for every Borel function $f$: $\lambda(f(A))=f(\lambda(A))$ $\forall A\in B_{sa}(\h)$.
\end{enumerate}
\end{theorem}

The connection of this theorem to physics stems from the fact that it can be used to prove that the following four statements cannot all be jointly true:

\begin{itemize}
\item[\textbf{QM}] (Quantum Mechanics) Every observable $\A$ can be associated with a self-adjoint operator $A$ on some Hilbert space $\h$. The result of a measurement of $\A$ is an element of the spectrum $\sigma(A)$ of $A$. Observables whose corresponding operators commute can be measured simultaneously. Moreover, if there is a functional relationship between the associated operators, this relation is preserved in the measurement results.
\item[\textbf{Re}] (Realism) Every observable $\A$ possesses a certain value $\lambda(\A)\in\mathbb{R}$ at all times.
\item[\textbf{FM}] (Faithful Measurement) A measurement of an observable $\A$ at a certain time reveals the value $\lambda(\A)$ possessed by that observable at that time.
\item[\textbf{CP}] (Correspondence Principle) There is a bijective correspondence between observables and self-adjoint operators.
\end{itemize}

Indeed, Re and CP together imply the existence of a function $\lambda:B_{sa}(\h)\to\mathbb{R}$ and FM and QM together imply that this function should satisfy 1 and 2. The Kochen-Specker Theorem can therefore be formulated in the following way:
\begin{equation}\label{QMKS}
	\mathrm{QM}\wedge \mathrm{Re}\wedge \mathrm{FM} \wedge \mathrm{CP}\to\bot
\end{equation}
Consequently, one is forced to give up at least one of these assumptions. QM states a part of quantum mechanics that is partly empirically testable and which seems hard to abandon in any theory that wishes to reproduce the experimental predictions of quantum mechanics. On the other hand, the conjunction of Re with FM is often seen as one of the great shortcomings not satisfied in orthodox quantum mechanics. Therefore, in any realist interpretation of quantum mechanics, endorsing Re and FM, one is likely to give up CP, which is the standard solution in all feasible hidden variable theories. Traditionally, it is taken that the negation of CP is likely to introduce contextuality; the idea that observables are not completely defined by their associated operator but only by also specifying the entire measuring context in which the observable is being measured. For example, Redhead stated that
\begin{quote}
``we might respond by [\ldots] rejecting Corr [CP in our text]. In other words, 
we now suppose that there are many different observables,  
corresponding in general to some particular operator $A$.'' \cite[p. 134]{Redhead87}
\end{quote}
We formulate non-contextuality in the following way:
\begin{itemize}
\item[\textbf{NC}] (Non-Contextuality) Every observable is uniquely defined by a self-adjoint operator.
\end{itemize}
which has a clear meaning in the present context, and avoids the notion of a measurement context. Although this notion may sound technical it is strongly related to more common definitions of contextuality that are formulated using triplets of self-adjoint operators $A, B, C$ such that $[A,B]=[A,C]=0$ but $[B,C]\neq0$. Then, a hidden variable theory is called non-contextual if for all such triplets the value assigned to $A$ does not depend on whether $A$ will be measured together $B$ or with $C$. From our NC it follows that there is only one observable corresponding to $A$ and therefore (by Re and FM) only one value assigned to this observable and operator. Thus the more common notion of non-contextuality follows from our NC with the aid of Re and FM. Strictly, the converse doesn't follow (making NC more restricting) since for NC it doesn't matter whether or not $B$ and $C$ commute, while for the common notion nothing is being said about the situation where $[B,C]=0$, but we will not make a fuzz about this detail.

However, another option for a hidden variable theory to violate CP is to not let every self-adjoint operator correspond to an observable. Indeed, CP is the conjunction of NC together with
\begin{itemize}
\item[\textbf{IP}] (Identification Principle) Every self-adjoint operator represents an observable.
\end{itemize}
Mathematically, one can look at these assumptions in the following way. If $\mathpzc{Obs}$ denotes the set of observables, then QM states that there exists a function $f:\mathpzc{Obs}\to B_{sa}(\h)$, NC states that this function is injective, and IP that this function is surjective.
The question then arises if it is actually possible to construct a hidden variable theory that is empirically equivalent with quantum mechanics, and satisfies Re, FM and NC, i.e. if rejecting IP is sufficient to construct a non-contextual hidden variable theory. We will show that this is possible by constructing such a theory.

\subsection{The finite-precision argument}\label{FPA}
The idea that not all self-adjoint operators represent observables is not new (e.g. \cite{Wigner52}). However, common proofs of the Kochen-Specker Theorem only assume a weakened version of IP; only a finite set of self-adjoint operators is selected for which it is assumed that they correspond to observables. Therefore, it is our task to motivate why not all of these operators that appear in any of these proofs should be identified with observables. The motivation will be well explained in the example of the original Kochen-Specker proof. In this proof one considers operators denoted $S_r^2$ which are associated with the squared spin of a spin-1 particle along some axis $r\in \mathbb{S}^2$, where $\mathbb{S}^2$ denotes the two-dimensional sphere in $\mathbb{R}^3$.

To circumvent the Kochen-Specker Theorem, we must assume that not for every $r\in \mathbb{S}^2$ the operator $S_r^2$ can be associated with an observable. We claim that it is sufficient if for every $r\in \mathbb{S}^2$ and every $\epsilon>0$ there exists an $r'\in \mathbb{S}^2$ such that $\|r-r'\|<\epsilon$ and $S_{r'}^2$ is associated with an observable. This mathematical requirement is explained in the following way. Suppose an experimenter wishes to measure the observable associated with $S_r^2$ for some $r\in \mathbb{S}^2$. Then it may be the case that this observable does not exist in our hidden variable model. For such objects we introduce the term \emph{pseudo observable}. The experimenter may not (and most likely cannot) know that this observable does not exist, but still can arrange the experimental setup. However, when he arranges his setup, he is only able to align the apparatus along the $r$-axis up to a finite precision bounded from below by a certain $\epsilon>0$. So instead of measuring the pseudo observable, it is assumed that the measurement reveals the value of some actual observable $\mathcal{S}_{r'}^2$ associated with the operator $S_{r'}^2$ for some $r'\in \mathbb{S}^2$ with $\|r-r'\|<\epsilon$. This is the finite-precision argument.

More generally, for empirical equivalence with quantum mechanics (up to arbitrary finite precision of measurements) it is sufficient to assume that there exists an injection $i:\mathpzc{Obs}\to B_{sa}(\h)$ from the set of observables to the set of self-adjoint operators such that $i(\mathpzc{Obs})$ is dense in $B_{sa}(\h)$ in the sense that $\forall A\in B_{sa}(\h),$ $\forall\epsilon>0$, $\exists \A'\in\mathpzc{Obs}$ such that $\|i(\A')-A\|<\epsilon$, where $\|.\|$ denotes the operator norm.

Since this specifies the setting for the rest of the article, we will assume further on that $\mathpzc{Obs}$ is specified as a subset of $B_{sa}(\h)$, and we will make no distinction in notation between observables and self-adjoint operators. Also, we note that the finite-precision argument merely provides an argument for selecting such a subset as the set of observables. But as soon as such a subset is constructed one could drop the argument and take it is an axiom that $\mathpzc{Obs}$ describes the set of observables. From this point of view, the MKC-models aren't models about the finite precision of measurement, which we think is a common misconception.

\subsection{Formulation of the MKC-models}\label{CK-sectie}

A requirement for a non-contextual hidden variable theory is that there exist sets of observables such that $\mathpzc{Obs}$ is dense in $B_{sa}(\h)$ that are also colorable in the Kochen-Specker sense (i.e., they satisfy 1. and 2. in Theorem \ref{KS-Theorem}). We will also assume that for any observable $A$ and Borel function $f$, $f(A)$ is again an observable. We thus aim to construct sets $\mathpzc{Obs}$ such that the set of hidden variables
\begin{equation}\label{variabelen}
	\Lambda:=\left\{\lambda:\mathpzc{Obs}\to\mathbb{R}\:;\:
	\substack{\forall A\in\mathpzc{Obs},\forall \text{ Borel function }f:\\ \lambda(A)\in\sigma(A),\; \lambda(f(A))=f(\lambda(A))}\right\}
\end{equation}
is not empty and $\mathpzc{Obs}$ is dense in $B_{sa}(\h)$. The hidden variables $\lambda\in\Lambda$ will also be called colorings of $\mathpzc{Obs}$.

That this is possible was proven in detail in \cite{CliftonKent99}. A formulation of this result is mandatory for the construction and understanding of the non-contextual MKC-models, but it requires some notations and definitions that we will formulate first.

For an orthonormal basis $\langle e_{i}\rangle_{i=1}^n$ we define
\begin{equation}
	\mathcal{P}_1\left(\langle e_{i}\rangle_{i=1}^n\right):=\{P_{e_1},\ldots,P_{e_2}\},
\end{equation}
where $P_{e_i}$ denotes the projection on the one-dimensional subspace spanned by $e_i$. Furthermore, let $\mathcal{P}(\langle e_{i}\rangle_{i=1}^n)$ denote the set of all projection operators that can be formed by making linear combinations of elements of $\mathcal{P}_1(\langle e_{i}\rangle_{i=1}^n)$ (including the zero operator) and set
\begin{equation}
	\mathfrak{A}\left(\langle e_{i}\rangle_{i=1}^n\right):=
	\left\{\sum_{i=1}^na_iP_{e_i}\:;\:a_i\in\mathbb{R}\forall i\in\{1,\ldots,n\}\right\},
\end{equation}
the Abelian C*-algebra generated by $\mathcal{P}_1\left(\langle e_{i}\rangle_{i=1}^n\right)$.

Now two orthonormal bases $\langle e_{i}\rangle_{i=1}^n$ and $\langle e'_{i}\rangle_{i=1}^n$ are called \emph{totally incompatible} if for all $P\in\mathcal{P}(\langle e_{i}\rangle_{i=1}^n)$ and $P'\in\mathcal{P}(\langle e'_{i}\rangle_{i=1}^n)$, $[P,P']=0$ if and only if $P\in\{0,\een\}$ or $P'\in\{0,\een\}$. In other words, $P\h\subset P'\h$ implies $P=0$ or $P'=\een$. Equivalently, two orthonormal bases are totally incompatible iff 
\begin{equation}
	\mathfrak{A}\left(\langle e_{i}\rangle_{i=1}^n\right)\cap
	\mathfrak{A}\left(\langle e'_{i}\rangle_{i=1}^n\right)=\mathbb{C}\een.
\end{equation}

\begin{theorem}[Clifton \& Kent 2001]\label{CKtheorem}
For every finite dimensional Hilbert space there exists a countable set of totally incompatible bases $\{\langle e_{1,i}\rangle_{i=1}^n,\langle e_{2,i}\rangle_{i=1}^n,\ldots\}$ such that
\begin{equation}
	\mathpzc{Obs}:=\bigcup_{k=1}^\infty\mathfrak{A}\left(\langle e_{k,i}\rangle_{i=1}^n\right)
\end{equation}
is dense in $B_{sa}(\h)$.
\end{theorem}

Colorings of this set are easy to construct. For any $f\in\{1,\ldots,n\}^{\mathbb{N}}$ we define $\lambda_f:\mathpzc{Obs}\to\mathbb{R}$ in the following way. If $A=c\een$, $\lambda_f(A):=c$. For every other $A$ there is exactly one $j\in\mathbb{N}$ such that $A\in\mathfrak{A}\left(\langle e_{j,i}\rangle_{i=1}^n\right)$ and we set 
\begin{equation}
	\lambda_f(A)=\lambda_f\left(\sum_{i=1}^na_{i}P_{e_{j,i}}\right):=a_{f(j)}.
\end{equation}
It is straight forward to show that this is indeed a coloring and that each possible coloring of $\mathpzc{Obs}$ is uniquely defined by an element of $\{1,\ldots,n\}^{\mathbb{N}}$ in this way, i.e. the set of all colorings (\ref{variabelen}) is isomorphic to $\{1,\ldots,n\}^{\mathbb{N}}$.\footnote{Proofs of these claims may be found in \cite[Lemma 3.4]{Hermens10}.} 

The obtained set $\Lambda$, of hidden variables, is rich enough to allow the definition of probability measures on it that obey the Born rule:

\begin{theorem}\label{Bornstelling}
For every density operator $\rho$ on the Hilbert space $\h$, there is a probability measure $\pee_{\rho}$ on $\Lambda$ such that
\begin{equation}\label{Born-regel}
	\mathbb{E}_{\rho}(A):=\int_{\Lambda}\lambda(A)\dee\pee_{\rho}(\lambda)=\Trace(\rho A)
\end{equation}
for every $A\in\mathpzc{Obs}$.
\end{theorem}

\noindent
\textit{Proof:}\footnote{This result was assumed to be trivial in \cite{CliftonKent99} and therefore no proof was given there. However, it is essential for the discussion and it seems not everyone was convinced that the MKC-models can in fact reproduce quantum statistics \cite{Cabello02}, \cite{Mermin99}. A more extensive discussion can be found in \cite[\S3.3 and \S3.5]{Hermens10}.}\hspace*{\fill}\\
In order to prove the existence of the probability measure $\pee_{\rho}$, $\Lambda\simeq \{1,\ldots,n\}^{\mathbb{N}}$ has to be turned into a measurable space first. For a finite sequence $t_1,\ldots,t_k$ of natural numbers (not necessarily in increasing order) and a sequence $B_1,
\ldots,B_k$ of subsets of $\{1,\ldots,n\}$, define 
\begin{multline}
	S(t_1,\ldots,t_k;B_1,\ldots,B_k):=\\
	\{\lambda_f\in\Lambda\:;\:f(t_1)\in B_1,\ldots,f(t_k)\in B_k\}.
\end{multline}
For a fixed sequence $t_1,\ldots,t_k$, let $\Sigma(t_1,\ldots,t_k)$ be the $\sigma$-algebra generated by all these sets.
Further, let $\Sigma$ be the smallest $\sigma$-algebra that contains all these $\sigma$-algebras.
Note that $\Sigma(t_1,\ldots,t_k)$ has in fact only finite many elements and that it is isomorphic (as a set) to the power set of $\{1,\ldots,n\}^{\{t_1,\ldots,t_k\}}$. Therefore, a probability measure on the space $(\Lambda,\Sigma(t_1,\ldots,t_k))$ is completely defined by its action on the sets that are equivalent to a singleton subset of $\{1,\ldots,n\}^{\{t_1,\ldots,t_k\}}$, i.e. the sets of the form
\begin{multline}
	 s(t_1,\ldots,t_k;j_1,\ldots,j_k):=\\
	 \{\lambda_f\in\Lambda\:;\:f(t_i)=j_i\text{ for }i=1,\ldots,k\}.
\end{multline}
\indent

Now let a density operator $\rho$ be given. For each finite sequence $t_1,\ldots,t_k$ define a probability measure $\pee_{\rho,t_1,\ldots,t_k}$ on the space $(\Lambda,\Sigma(t_1,\ldots,t_k))$ by
\begin{equation}
\pee_{\rho,t_1,\ldots,t_k}[s(t_1,\ldots,t_k;j_1,\ldots,j_k)]
	:=
	\prod_{i=1}^k\Trace\left(\rho P_{e_{t_i,j_i}}\right),
\end{equation}
for each sequence $j_1,\ldots,j_k$ in $\{1,\ldots,n\}$.

It is easy to see that these probability measures satisfy the following consistency criteria:
\begin{enumerate}
\item For any finite sequence $t_1,\ldots,t_k$ and all permutations $(t_1',\ldots,t_k')=(t_{\pi(1)},\ldots,t_{\pi(k)})$ one has 
\begin{multline}
	\pee_{\rho,t'_1,\ldots,t'_k}[s(t'_1,\ldots,t'_k;j_{\pi(1)},\ldots,j_{\pi(k)})]=\\
	\pee_{\rho,t_1,\ldots,t_k}[s(t_1,\ldots,t_k;j_1,\ldots,j_k)].
\end{multline}
\item For each finite sequence $t_1,\ldots,t_k,t_{k+1}$, the measure $\pee_{\rho,t_1,\ldots,t_k,t_{k+1}}$ acts as $\pee_{\rho,t_1,\ldots,t_k}$ on every set in $\Sigma(t_1,\ldots,t_k)$, i.e. $\pee_{\rho,t_1,\ldots,t_k,t_{k+1}}(S)=\pee_{\rho,t_1,\ldots,t_k}(S)$ for all $S\in\Sigma(t_1,\ldots,t_k)$. 
\end{enumerate}
Then, according to the Kolmogorov's extension theorem (see for example Theorem 10.1 in \cite{Bhat07}), there exists a probability measure $\pee_{\rho}$ on the space $(\Lambda,\Sigma)$ such that $\pee_{\rho}$ acts as $\pee_{\rho,t_1,\ldots,t_k}$ on the sets in $\Sigma(t_1,\ldots,t_k)$, for each finite sequence $t_1,\ldots,t_k$.

The only thing left to show is that $\pee_{\rho}$ satisfies (\ref{Born-regel}). For one-dimensional projections this follows almost immediately:
\begin{equation}
\begin{split}
	\mathbb{E}_{\rho}(P_{e_{m,l}})&=\pee_{\rho}[P_{e_{m,l}}=1]\\
	&=\pee_{\rho}(\{\lambda_f\in\Lambda\:;\:f(l)=m\})\\
	&=\pee_{\rho,l}(\{\lambda_f\in\Lambda\:;\:f(l)=m\})=\Trace(\rho P_{e_{m,l}}).
\end{split}
\end{equation}
And also for any multiple of the unit operator (\ref{Born-regel}) follows immediately since
\begin{equation}
\begin{split}
	\mathbb{E}_{\rho}(c\een)&=\int_{\Lambda}\lambda(c\een)\dee\pee_{\rho}(\lambda)
	=c\int_{\Lambda}\dee\pee_{\rho}(\lambda)\\
	&=c
	=\Trace(\rho c\een).
\end{split}
\end{equation}
For all the other observables $A\in\mathpzc{Obs}$, there is exactly one $m\in\mathbb{N}$ with $A\in\mathfrak{A}\left(\langle e_{m,i}\rangle_{i=1}^n\right)$. Therefore, 
\begin{equation}
	\mathbb{E}_{\rho}(A)
	=
	\sum_{j=1}^na_{j}\mathbb{E}_{\rho}(P_{m,j})=
	\sum_{j=1}^na_{j}\Trace(\rho P_{m,j})
	=\Trace(\rho A).
\end{equation}
This completes the proof.
\hfill $\square$\\[0.5ex]

Any hidden variable model that uses this set up may be called an MKC-model. There are of course still many ways to complete this to form a complete theory, in the sense that no interpretation for the hidden variables and the probability measures has been specified yet, and no time-evolution of the state of the system has been defined. In the next section we propose a specific MKC-model that is complete in this sense, and for which we claim that it is completely empirically equivalent to quantum mechanics.

\subsection{The Hedgehog Interpretation of Quantum Mechanics}\label{HHint}

Although Barrett and Kent claimed that the MKC-models show the existence of non-contextual hidden variable theories \cite{Barrett-Kent04}, they did not bother to proof this explicitly. Overmore, it was not clear why some of the objections against these models were incorrect or irrelevant (although they did make some terrific efforts in this direction). Most noticeably are the objections due to Appleby \cite{Appleby00}, \cite{Appleby01}, \cite{Appleby02}, \cite{Appleby05}. He not only objected that no time-evolution for the MKC-models were specified ever, but also showed why there are difficulties with respect to this issue. Here, we will show that these difficulties can be resolved.

The hidden variables in the MKC-models have a seemingly paradoxical character. On one hand, their definition is motivated by considering the finite precision of measurements, but on the other hand, they have been accused for being unphysical using finite precision arguments as well \cite{Appleby00}, \cite{Mermin99}. The reason is that they are likely to assign totally different values to observables that are very close to each other in $\mathpzc{Obs}$, a result that was proven by Appleby for the squared spins of a spin-1 particle \cite{Appleby05}. Explicitly, he showed the following:
\begin{theorem}\label{appelstelling}
	Let $\mathpzc{Obs}\subset B_{sa}(\mathbb{C}^3)$ be a dense colorable subset and let $\lambda$ be a coloring of this set. Then there exists an open set $U_{\lambda}$ in the 2-sphere $\mathbb{S}^2$ such that $\lambda$ restricted to $\{S_r^2\in\mathpzc{Obs}\:;\:r\in U_{\lambda}\}$ is discontinuous in every point.
\end{theorem}

At a first glance, this result doesn't seem to give rise to any problems. Any coloring is trivially discontinuous at some points for it can only assign the values 0 and 1 on the set $\{S_r^2\in\mathpzc{Obs}\}$, and both values must be obtained. So any objection based on Theorem \ref{appelstelling} should exploit the fact that the discontinuities are \emph{dense}.\footnote{This point was already discussed in \cite[$\S$3.1]{Barrett-Kent04}. But no satisfactory solution was obtained for the objections we will raise.} We will illustrate a possible objection and construct a specific MKC-model that circumvents this objection. This model will be termed the `hedgehog interpretation' of quantum mechanics.

Consider a spin-1 particle in a non-contextual hidden variable state $\lambda$, and let $U_{\lambda}$ be as in Theorem \ref{appelstelling}. Now suppose one wishes to subsequently measure the (possibly pseudo) observable $S_r^2$ for some $r\in U_{\lambda}$. Quantum mechanics predicts that the result of every measurement will be the same. However, in the MKC-model this seems unlikely for there is no reason to assume that the subsequent measurements are actually measurements of the same observable; even the slightest variation in the measurement setting is likely to change the obtained measurement result. 

A first possible solution to this problem, which we do not find satisfactory, may be to introduce an extra hidden variable that serves as a memory of which measurements have been performed on the system. The actual observable then being measured depends both on the setting of the measurement apparatus and the previous obtained measurement results. Although conspirational, this scheme may seem to work as long as the actual observable being measured lies within the finite precision range of the observable that was intended to be measured. However, the notion of `finite precision range' is not well-defined and an experimenter may set out to `trick nature' in this scheme. Indeed, the experimenter may not be setting out to subsequently measure the same observable, but to subsequently measure different observables within some small neighborhood. The system thus would have to know if the small derivations in the setting of the measurement apparatus are due to the limited precision of measurement, or that they are intentional. A scheme that is too conspirational for this author.

Instead, we adopt an easier solution to the problem by actually copying the solution von Neumann already introduced for quantum mechanics \cite[Ch. III, $\S$3]{Neumann55}. Indeed, if one assumes that the measurement changes the state $\lambda$ into the state $\lambda'$ --where $\lambda'$ is randomly selected according to the measure $\pee_{\rho'}$ and $\rho'$ is the projected quantum mechanical state-- subsequent measurements are completely in accordance with quantum mechanical predictions. But there's no reason to stop here. It is also desirable that the evolution of the hidden variable state is in accordance with all the statistical predictions of quantum mechanics and thus the unitary evolution of the quantum state should also be incorporated in the model . It then seems natural to describe the evolution of the hidden variable state $\lambda(t)$ as a stochastic process such that for every $t\in\mathbb{R}$ and measurable set $\Delta\subset\Lambda$ one has
\begin{equation}
	\pee[\lambda(t)\in\Delta]=\pee_{\rho(t)}(\Delta),
\end{equation}
where the evolution of $\rho(t)$ is given by the postulates of quantum mechanics (including the projection postulate).

The MKC-model specified above presents itself as an interpretation of orthodox quantum mechanics, where all observables have a non-contextual value at all times. The set of observables in this interpretation coincides with a `spiked' subset of the set of all self-adjoint operators, hence the term `hedgehog interpretation'. By construction it is empirically indistinguishable from quantum mechanics and therefore proves the existence of a non-contextual hidden variable theory. In other words, although
\begin{equation}
	\mathrm{QM}\wedge \mathrm{Re}\wedge \mathrm{FM} \wedge \mathrm{NC} \wedge\mathrm{IP}\to\bot
\end{equation}
is true, giving up either NC or IP is sufficient to prevent a contradiction, and explicit models that show this can be constructed (as we have shown here for only giving up IP). One may note that from this point of view, the Kochen-Specker Theorem is not nullified by the finite precision argument (motivating the rejection of IP) any more than it already was by Bell's Bohrian insight in \cite{Bell66} where he argued that certain demands (like NC) ``are seen to be quite unreasonable when one remembers with Bohr `the impossibility of any sharp distinction between the behavior of atomic objects and the interaction with the measuring instruments which serve to define the conditions under which the phenomena appear.' \cite[p. 210]{Bohr49}''.\footnote{Historically, somewhere something must have went wrong with the interpretation of the Kochen-Specker Theorem when the consensus became that Bell's argument against the original theorem (a variation of (\ref{QMKS})) was the only possible argument against it, i.e. that giving up NC is the only way to violate CP. So although the theorem has now become famous as the proof of the impossibility of non-contextuality (which it doesn't prove), it originally wasn't set up to prove this but merely to show that a hidden variable model for quantum mechanics that satisfies certain seemingly plausible requirements cannot be constructed (which it does prove).} 

One may of course argue that a hidden variable theory should at least solve the measurement problem, and that a hidden variable theory with a projection postulate is not a hidden variable theory at all. We however believe that the problem of realism (in the sense of Re) and the measurement problem may be independent and should thus be treated independently, i.e. the measurement problem is not part of the present discussion. There is however still the question whether a satisfactory MKC-model may exist that circumvents the use of the projection postulate, for example, in the form of a spontaneous collapse theory. Indeed, the MKC-models show a lot of flexibility when it comes to possible time-evolutions, and it is not unlikely that they can also serve as explicit toy models in the investigation of other problems in the foundations of quantum mechanics. For example, the models are flexible enough to allow versions that violate the so-called Tsirelson bound for the CHSH-inequality \cite{Tsirelson80}, \cite{Clauser69}.\footnote{A maximal violation of the inequality can easily be achieved for a pair of spin-1 particles by introducing a specific time-evolution. First, consider for one particle the operator $Q_r=2S_r-\een$ and let $\psi_r$ denote its eigenstate for the eigenvalue -1. For the composite system we introduce the pure quantum states $\rho_{rs}=(\psi_r\otimes\psi_s)(\psi_r\otimes\psi_s)^*$. Now we define the operators $A=Q_x\otimes\een$, $A'=Q_y\otimes\een$, $B=\een\otimes Q_x$ and $B'=\een\otimes Q_y$. Let the initial state of the system $\lambda$ be chosen according to the measure $\pee_{\rho_{zz}}$. A measurement of any of the (pseudo) observables $A,A',B,B'$ will yield the result 1. Now we stipulate that a measurement of $A$ or $B$ doesn't change the state, but a measurement of $A'$ or $B'$ does cause it to change to a new state $\lambda'$ selected according to the measure $\pee_{\rho_{yy}}$. Then we find that $|\mathbb{E}(AB)+\mathbb{E}(AB')+\mathbb{E}(A'B)-\mathbb{E}(A'B')|=4$. It is not clear if such an easy example is possible for a pair of spin-$\tfrac{1}{2}$ particles, nor if it is possible to obtain the same violation using a more natural time-evolution. An other possibility for violating the Tsirelson bound may be by also allowing probability measures other that the ones defined in Theorem \ref{Bornstelling}, i.e. measures that don't obey the Born-rule.} In the remainder of this paper, we will use them to improve the discussion on the possibility of experimental tests for contextuality.

\section{Operational definitions of contextuality}\label{operational-sectie}
\subsection{Introduction}
Thus far we have mainly focused on one notion of contextuality and one may argue that this isn't general enough for the present discussion. In \cite{Spekkens05}, Spekkens proposed a generalization that, allegedly, applies to all operational theories, arbitrary experimental procedures and a broad class of ontological models of quantum theory, namely:
\begin{quote}
	``A noncontextual ontological model of an operational theory is one wherein if two experimental procedures are operationally equivalent, then they have equivalent representations in the ontological model.'' \cite{Spekkens05}
\end{quote}
Obviously, there is a lot of terminology in this definition which has to be defined before we can investigate it. However, a few notes of critique are already at place. First of all, this notion of non-contextuality isn't really a general notion that applies to all operational theories, but rather, it is a notion that is defined for all operational theories but that may have a different meaning for different operational theories. Indeed, the precise notion of operational equivalence can only be defined within the theory. It is therefore not a metaphysical notion but a rather technical one. 

This brings us to a second point. If one confines oneself only to quantum mechanics, the definition only applies to ontological models that can be related to the operational approach of quantum mechanics. We believe the operational approach is clouded by the instrumentalist philosophy, and it can be taken as one of the aims of constructing hidden variable models to get rid of this philosophy. To elaborate, in \cite{Spekkens05} an operational theory is a theory in which preparations, transformations and measurement procedures are the elementary concepts, and we see no need why these concepts should play a role in a hidden variable model per se. Of course there is still the question of empirical equivalence between the operational approach and the hidden variable approach, but in our opinion this doesn't imply that it should be possible that all operational concepts can be translated in the language of the hidden variable approach. In contrast, the notion of outcomes of measurements of observables associated with self-adjoint operators (upon which NC is based), is one that clearly is necessary in any approach to quantum mechanics.

Let us introduce the necessary terminology. In an operational theory, an experimental procedure is described by a preparation $P$, a transformation $T$ and a measurement $M$. When all three components are specified, they give rise to a probability measure $\pee[\;.\;|P,T,M]$ on the set of possible outcomes for the experiment. Two preparations $P$ and $P'$ are equivalent iff $\pee[\;.\;|P,T,M]=\pee[\;.\;|P',T,M]$ for all $T$ and $M$. Similarly, two transformations $T$ and $T'$ are equivalent iff $\pee[\;.\;|P,T,M]=\pee[\;.\;|P,T',M]$ for all $P$ and $M$ and two measurements are equivalent iff $\pee[\;.\;|P,T,M]=\pee[\;.\;|P,T,M']$ for all $P$ and $T$.
Two experimental procedures are (operationally) equivalent if their preparation, transformation and measurement are equivalent. 

A representation of an (operationally defined) experimental procedure in the ontological model consists of a representation of the preparation, the transformation and the measurement. Because the experimental procedure is described by three components, one can also split the notion of non-contextuality in three components. That is, an ontological model is called preparation non-contextual if operationally equivalent preparations have equivalent representations in the ontological model. Similarly, one can define transformation non-contextuality and measurement non-contextuality. Thus an ontological model is non-contextual iff it is preparation non-contextual, transformation non-contextual and measurement non-contextual. 

\subsection{Preparation non-contextuality}
A representation of the preparation $P$ is a probability measure $\pee_P$ on the set of ontological states $\Lambda$. Thus a preparation $P$ prepares the system in the ontological state $\lambda$ with a probability according to this measure $\pee_P$. Two representations $\pee_P$ and $\pee_P'$ are equivalent iff they are equal. So an ontological model is preparation non-contextual if every representation of a preparation $P$ is completely specified by the equivalence class of preparations to which $P$ belongs. In short,
\begin{equation}\label{prepNC}
	\forall T, M \pee[\;.\;|P,T,M]=\pee[\;.\;|P',T,M]\:\alsdan\:\pee_P=\pee_{P'}.
\end{equation}
In quantum mechanics every equivalence class of preparations can be identified with a density operator $\rho$. So an ontological model of quantum mechanics is preparation non-contextual if every representation of a preparation is of the form $\pee_{\rho}$, where $\rho$ is the density operator associated with the preparation in the operational approach to quantum mechanics. 

In the MKC-models probability measures are in fact specified by the density operator rather than by some notion of a preparation, and it therefore seems that these models are trivially preparation non-contextual. On the other hand, in \cite{Spekkens05} it was proven that every ontological model must be preparation non-contextual. Since the proof takes place in the setting of the Hilbert space $\mathbb{C}^2$, an argument against it based on the finite precision argument isn't possible. Technically, this is because every set of orthonormal bases of $\mathbb{C}^2$ is totally incompatible, and therefore every set of self-adjoint operators is a possible set of observables in an MKC-model. So the discrepancy must be sought elsewhere.

It turns out that the MKC-models do not satisfy one of the criteria every preparation non-contextual theory should obey according to \cite{Spekkens05}: every convex combination of preparation procedures must be represented within the ontological model by a convex sum of the associated probability measures. Indeed, suppose $\rho,\rho'$ and $\rho''$ are three distinct density operators such that $\rho=p\rho'+(1-p)\rho''$ for some $p\in[0,1]$, then, in general (in the MKC-models), one does not have that $\pee_{\rho}=p\pee_{\rho'}+(1-p)\pee_{\rho''}$. For example, let $P_1$ and $P_2$ be two non-commuting one-dimensional projections and set $\rho=\tfrac{1}{2}P_1+\tfrac{1}{2}P_2$, then
\begin{equation}
\begin{split}
	\pee_{\rho}[P_1=1,P_2=1]
	&=\Trace(\rho P_1)\Trace(\rho P_2)\\
	&=\left(\frac{1}{2}+\frac{1}{2}\Trace(P_1P_2)\right)^2,
\end{split}
\end{equation}	
which in general is not equal to
\begin{multline}
  \frac{1}{2}\pee_{P_1}[P_1=1,P_2=1]+\frac{1}{2}\pee_{P_2}[P_1=1,P_2=1]=\\
	\frac{1}{2}\Trace(P_1)\Trace(P_1 P_2)+\frac{1}{2}\Trace(P_2 P_1)\Trace( P_2)=\\
	\Trace(P_1 P_2).
\end{multline}
This is despite the fact that 
\begin{equation}
\begin{gathered}
  \pee_{\rho}[A\in\Delta]=p\pee_{\rho'}[A\in\Delta]+(1-p)\pee_{\rho''}[A\in\Delta],\\
  \quad\forall A\in\mathpzc{Obs},\Delta\subset\sigma(A),
\end{gathered}
\end{equation}
which means that the two measures are empirically equivalent for all quantum mechanical admissible measurements, but not for subsets that do not correspond to possible observations. Of course, this relies on the assumption that observables corresponding to non-commuting operators are not jointly measurable. If one rejects this assumption, the two measures $\pee_{\rho}$ and $p\pee_{\rho'}+(1-p)\pee_{\rho''}$ are empirically distinguishable (e.g. by simultaneous measurements of $P_1$ and $P_2$).

So, if one accepts this line of reasoning, the MKC-models are preparation non-contextual. There are a few questions relevant at this point. First, can a convex combination of preparations be associated with a preparation? To this question we answer in the same way as Spekkens would: of course. If one can prepare a system in the state $\rho'$ and in the state $\rho''$ and one is able to produce a random variable $X$ that gives the value 0 with probability $p$ and the value 1 with probability $1-p$, one can let the outcome of this experiment decide whether one prepares the system in the state $\rho'$ (upon finding 0), or in the state $\rho''$ (upon finding 1). Then, for all possible measurements, the system is prepared in such a way as if it were prepared in the state $\rho=p\rho'+(1-p)\rho''$. The question then is whether or not this preparation procedure is operationally equivalent to a preparation in the state $\rho$. If one forgets about the preparation procedure the answer would be yes since in that case no future measurement can distinguish between the two preparation procedures. However, there is a clear physical difference, even operationally, because one can distinguish by simply measuring the value of $X$. On a more abstract level, we think that the equivalence of two preparation procedures is only well defined if both preparation procedures are entirely defined for the same system under consideration. So if one wishes to study a certain system, all possible preparation procedures must be specified in terms of this system.

Even if one doesn't accept this argument against the proof of the preparation contextuality of quantum mechanics in \cite{Spekkens05}, there is still the question of the harmfulness of preparation contextuality. In our opinion, it isn't that weird that two physically distinct preparations would lead to distinct preparations of the ontological state. The only thing that is seemingly weird about the situation would be that, once one forgets about the preparation, there may be no possible measurement that recovers information about the preparation. But we don't think this is typically non-classical; throwing away information may lead to the impossibility of retrieving that information even classically. In fact, in section \ref{Spekkenstest} we will show that preparation contextuality, as it is understood in \cite{Spekkens05}, also appears in classical systems.  

\subsection{Transformation non-contextuality}
Our discussion on the notion of transformation non-contextuality would be quite similar and so we wish to skip the most of it. However, there are some interesting additional remarks to make. Spekkens requires that every possible transformation $T$ is represented by a transition matrix $\Gamma_{T}(\lambda,\lambda')$ which would describe the probability of the transition from the ontic state $\lambda'$ to $\lambda$ given the transformation $T$. In the MKC-models no such object was defined. Mostly because we felt no need for it, but also because we were a bit sloppy. Thus far, the hedgehog interpretation is our only MKC-model candidate for this discussion, since it is the only one with a time evolution. Although we introduced the evolution of the state $\lambda$ as a stochastic process, the process hasn't been completely defined mathematically. Our only requirement was that 
\begin{equation}\label{nogeen}
	\pee[\lambda(t)\in\Delta]=\pee_{\rho(t)}[\Delta],
\end{equation} 
but $\pee$ wasn't defined, and there is no unique way for doing this. Formally, $\pee$ is a measure on the space $\Lambda^{\mathbb{R}}$ and one needs to define its value on every measurable subset. For the investigation of transformation non-contextuality we at least require to define the values for $\pee[\lambda(t)\in\Delta,\lambda(t')\in\Delta']$, and they don't follow uniquely from (\ref{nogeen}). The easiest way to introduce such a measure is by assuming that $\lambda(t)$ and $\lambda(t')$ are stochastically independent whenever $t\neq t'$. One then simply has that 
\begin{equation}
	\pee[\lambda(t)\in\Delta,\lambda(t')\in\Delta'] =\pee_{\rho(t)}[\Delta]\pee_{\rho(t')}[\Delta'].
\end{equation}
In other words, knowing the precise state $\lambda$ at a certain point in time doesn't help one at all in making predictions about the future state. 

This discussion isn't very illuminating on the subject of transformation non-contextuality, but it does demonstrate that the assumption that transition matrices $\Gamma_T$ should be defined in an ontological model isn't a very trivial assumption. In fact, we are still not at the point where we have defined such objects for the hedgehog interpretation. That would involve a lot more mathematical discussion that isn't interesting for the present discussion. For now, we have shown that any transformation in the hedgehog interpretation is completely specified by the transformation of the quantum mechanical state $\rho$. This sufficiently shows that the hedgehog interpretation is transformation non-contextual if one accepts that equivalence classes of transformations can be specified by operators acting on the quantum mechanical state. 

However, in \cite{Spekkens05} it is also claimed that every ontological model must be transformation contextual. The puzzling assumption in the proof of this claim is similar to the one we criticized for preparation non-contextuality. Namely, it is assumed that convex combinations of transformations are represented by convex combinations of transition matrices. This assumption gives rise to the same discussion that was given for preparations.

\subsection{Measurement non-contextuality}
Finally, we consider the notion of measurement non-contextuality. A representation of a measurement $M$ in an ontological model is a function $f_M$, which assigns to each $\lambda\in\Lambda$ a probability distribution on the set of possible outcomes for the experiment. In the MKC-models, every measurement is taken to be the measurement of an observable $A$ which is described quantum mechanically by the projection valued measure (PVM) $\mu_A$. The function $f_A$ simply assigns to each $\lambda$ the value $\lambda(A)$ which denotes the outcome of the measurement. In this sense, the MKC-models are deterministic (in \cite{Spekkens05} this is called `outcome determinism'). Because in this sense two measurements are operationally equivalent iff they are associated with the same observable, the MKC-models are measurement non-contextual. 

Of course, in \cite{Spekkens05} a proof is presented that shows the necessity of measurement contextuality for every ontological model. Although it requires that the ontological model also describes positive operator valued measures (POVM), this is not where our main concern lies. In fact, MKC-models that also incorporate POVM's may also be defined (c.f. \cite{CliftonKent99}, \cite{Barrett-Kent04}). No, our problem is that again auxiliary systems and measurements thereupon are introduced, and then one forgets about this procedure in the description of the measurement process once the result has been obtained. Indeed, although the probability distributions over the possible outcomes of measurements are the same in both systems, it is our opinion that the measurement of the spin along some axis of a spin-$\tfrac{1}{2}$ particle prepared in the state $\rho=\tfrac{1}{2}\een$ isn't operationally equivalent to the flipping of a fair coin and pretending the result came from a measurement on the spin-$\tfrac{1}{2}$ particle (whereas Spekkens claims that it is).    
  
All in all we believe that the notions of contextuality defined in \cite{Spekkens05} rely too much on the operational approach as is made clearer by our discussion. We have also seen that the notion of `operational equivalence' can become quite ambiguous when discussing ontological models. The operational notions of non-contextuality are also not that interesting metaphysically; they require that physically distinct situations, which by some bending of definitions become operationally equivalent, must have indistinct representations in ontological models. It is this strong and unsatisfying feature that allows Spekkens to proudly construct proofs in a Hilbert space of dimension two, and claiming that therefore they are ``stronger than traditional proofs of contextuality'', which can only be constructed in Hilbert spaces of dimension three and higher. It is clear that in this case the stronger result relies on much stronger assumptions.

\section{Experimental tests for contextuality}\label{experimental-sectie}

\subsection{Introduction}\label{paradox}

The Kochen-Specker Theorem and Bell's inequalities probably are the most celebrated results in the foundations of quantum mechanics of the past fifty years. While the latter has been the subject for experimental tests for over thirty years, the discussion on experimental tests for the Kochen-Specker Theorem had to wait for a boost until the turn of the century (e.g. \cite{Bartosik09}, \cite{Cabello08}, \cite{Cabello10}, \cite{Guhne10}, \cite{Huang03}, \cite{Kirchmair09}, \cite{Klyachko08}, \cite{Liu09}, \cite{Michler00},  \cite{Moussa10}, \cite{Pan10}, \cite{Plastino10}, \cite{Simon00}, \cite{Spekkens09}). In accordance with the general idea that the Kochen-Specker Theorem excludes non-contextual hidden variable theories, these experimental tests are also often referred to as experimental tests of contextuality.

In principle, we are suspicious of the idea of an experimental test of contextuality.
In our opinion, a test to prove contextuality would require a simultaneous measurement of a single observable in two different measuring contexts. These measuring contexts need not even be incompatible in the quantum-mechanical sense. The paradox is of course that, if the two measurements are distinguishable, one cannot maintain the idea that the observable is the same in both experiments, whereas, if they are indistinguishable, there is only one measurement of one observable and not two. Roughly, this is why we believe that experimental tests of contextuality are impossible, but a more careful investigation is still required, which we will provide in the remainder of this article.

The mentioned paradox would supposedly be circumvented by deriving statistical laws that are to be true in every non-contextual hidden variable theory, but violated in experiment and by quantum mechanics. However, the derivation of every of these laws require more than the assumption of Re, FM and NC; our minimal assumptions for a non-contextual hidden variable theory. Although these laws come in a huge variety, they basically make the same implicit auxiliary assumptions. One of the more famous \cite{Cabello08} is based on a proof of the Kochen-Specker Theorem known as the Mermin-Peres square for two spin-$\tfrac{1}{2}$ particles (e.g. \cite{Mermin93}):
\begin{equation*}
\xymatrixrowsep{1.4pc}
\xymatrixcolsep{1.4pc}
\xymatrix{
		A_{11}=\sigma_x\otimes\een \ar@{-}[r]\ar@{-}[d] &
		A_{12}=\een\otimes\sigma_x \ar@{-}[r]\ar@{-}[d] &
		A_{13}=\sigma_x\otimes\sigma_x \ar@{-}[d] \\ 
		A_{21}=\een\otimes\sigma_y \ar@{-}[r]\ar@{-}[d] &
		A_{22}=\sigma_y\otimes\een \ar@{-}[r]\ar@{-}[d] &
		A_{23}=\sigma_y\otimes\sigma_y \ar@{-}[d] \\ 
		A_{31}=\sigma_x\otimes\sigma_y \ar@{-}[r] &
		A_{32}=\sigma_y\otimes\sigma_x \ar@{-}[r] &
		A_{33}=\sigma_z\otimes\sigma_z}
\end{equation*} 
Here $\sigma_x,\sigma_y$ and $\sigma_z$ denote the Pauli spin matrices for a single spin-$\tfrac{1}{2}$ particle, and $\een$ is the unit operator on $\mathbb{C}^2$. Every row and every column in this square consists of three commuting operators, which, according to quantum mechanics, can be measured simultaneously. One can therefore introduce the quantum observables $R_{k}=A_{k1}A_{k2}A_{k3}$ and $C_k=A_{1k}A_{2k}A_{3k}$, which are identified with measuring the three operators in the specified row/column simultaneously and then taking their product. Quantum mechanics predicts that the result of a measurement of $A_{ij}$ equals $\pm1$, and the result of $R_k$ and $C_k$ is always 1, except for $C_3$, which will be -1. However, one cannot assign values to all $A_{ij}$ independent of them appearing in a row or column such that all these rules are satisfied. In other words, the assignment of definite values to these operators requires the introduction of contextuality (in the sense of NC). 

This result is of course not experimentally testable since one cannot measure any $A_{ij}$ simultaneously both in the context of the row in which it appears and the column in which it appears to check if the values are identical because the experimental contexts are mutually exclusive. But one can derive (cf. \cite{Cabello08}) the following statistical inequality for all possible non-contextual value assignments to all $A_{ij}$:
\begin{equation}\label{Cabineq}
	\mathbb{E}(R_1)+\mathbb{E}(R_2)+\mathbb{E}(R_3)+\mathbb{E}(C_1)+\mathbb{E}(C_2)-\mathbb{E}(C_3)\leq4,
\end{equation}
whereas quantum mechanics predicts the value 6 for the left-hand side for all possible states of the system. 

It is somewhat astonishing that so many authors believe that because (\ref{Cabineq}) is an experimentally testable inequality even if the experiments only have a finite precision (and we agree with this), it somehow annihilates the finite-precision argument against contextuality. However, the finite-precision argument still holds, and it rules out the derivation of (\ref{Cabineq}) on the same grounds as that it ruled out the old argument based on the Mermin-Peres square: one does not have to accept that all the operators appearing in the square correspond to observables. In fact, by the construction of Theorem \ref{CKtheorem}, if $A_{11},A_{12}$ and $A_{13}$ are identified with observables, all the other $A_{ij}$ are pseudo observables, and so are $R_2,R_3,C_1,C_2$ and $C_3$. Indeed, the starting assumption of the entire derivation of (\ref{Cabineq}) is a weakened version of IP  (c.f. section \ref{FPA}) that is rejected in the MKC-models.

\subsection{The behavior of MKC-models in experimental tests of contextuality}
It is of interest to see how the MKC-models \emph{can} manage to violate (\ref{Cabineq}). It is also non-trivial; the fact that the derivation of (\ref{Cabineq}) uses assumptions that aren't true in these models, doesn't imply that these models can actually violate (\ref{Cabineq}). For this we distinguish two possible ways of testing (\ref{Cabineq}). 

First we consider single measurements carried out on systems that are part of an ensemble of systems prepared in the same quantum state $\rho$. That is, on each system in the ensemble only one measurement is performed corresponding to either a row or column in the Mermin-Peres square.\footnote{By a single measurement on a row/column, we mean a simultaneous measurement of all the observables in the row/column and then taking the product of the obtained values.} In every MKC-model, each of these systems is in a state $\lambda$ selected according to the measure $\pee_{\rho}$. Then, by Theorem \ref{Bornstelling}, one will find the value 1 for each of the expectation values in (\ref{Cabineq}) with exception of the last term, for which one will find the value -1, resulting in a violation of the inequality. 

Next, consider the situation where the inequality is tested by performing subsequent measurements on the same system. One may follow the same line of reasoning and come to the conclusion that (\ref{Cabineq}) is still violated. However, this test also provides a way to discriminate between several MKC-models. For example, if a measurement of $R_1$ is followed by a measurement of $C_1$, not every MKC-model predicts that the value obtained for the (possibly pseudo) observable $A_{11}$ will be the same in both measurements. Obtaining different values would of course not prove contextuality; in the MKC-model the actual observable measured is not the same one in both experiments. But it does show a discrepancy with quantum mechanics which predicts that the obtained value for $A_{11}$ \emph{will} be the same in both situations. Of course, the hedgehog interpretation of section \ref{HHint} is one of the exceptions since in that model the evolution of the state $\lambda$ is changed by the first measurement in such a way that the obtained value for $A_{11}$ is the same in both measurements.

If one turns the argument around one again recognizes the flexibility of the MKC-models; even if experimental tests with subsequent measurements would disprove quantum mechanics and the hedgehog interpretation (e.g. one does find varying values for $A_{11}$), there is still an MKC-model with a time evolution that conforms with these measurements.  

It should be noted that for the Mermin-Peres square an other non-contextual model that is not of the MKC-type has been proposed by La Cour in \cite{LaCour09}. This model \emph{does} satisfy NC and IP for the nine observables in the square but it can work because it exploits a specific loophole in the proof of the Kochen-Specker Theorem. QM together with Re and FM requires that when observables with some functional relationship are being measured, the measurement results obey the same functional relations. However, this doesn't imply that the values possessed by these observables should obey these relations at all times. And there are many ways to assign values to all observables if one doesn't require these functional relationships between observables. La Cour exploits this loophole by stipulating that before a measurement the measurement apparatus interacts with the system in such a way that the values of the observables that can be measured are altered such that they do obey the required functional relationships. Of course this can be seen as a violation of FM since the measurement result obtained is not the value possessed by the observable when the measurement process is initiated, but the value after the process has occurred. It remains to be investigated if such a scheme can be generalized such that one can construct non-contextual hidden variable models for all possible quantum systems. 

Meanwhile, it should be clear that the MKC-models can violate other proposed inequalities for non-contextuality (like the one proposed in \cite{Klyachko08}) that are similar to (\ref{Cabineq}) (in a sense as explained in \cite{Cabello10-2}), in the same manner as described above. Consequently, the result in \cite{Cabello10-3} which is a critique on the original model of Meyer doesn't apply to the more advanced MKC-models. In fact, the model of Meyer --where one only considers squared spin observables along an axis specified by unit vectors with rational components-- isn't an MKC-model in the sense of how we defined them here. Specifically, the notion of total incompatibility, which requires that each observable can only appear in one measurement context (c.f. section \ref{CK-sectie}), isn't satisfied in Meyer's model. 


\subsection{Non-locality and `existential contextuality'}

Most of the recent articles on contextuality pay minor to no attention to the finite precision argument. The recent \cite{Cabello10} is an exception, and it is claimed in that paper that the finite-precision `loophole' can be closed. Astonishingly, this is done by introducing a locality argument. Indeed, one now considers measurements on spatially separated particles and they argue that
\begin{quote}
	 ``[s]patial separation provides a physical basis to the assumption that both measurements [on the spatially separated particles] are not only approximately but perfectly compatible.'' \cite{Cabello10}
\end{quote}
In light of the Mermin-Peres square, they argue that one may assume that $A_{11}$ is an observable and that both $A_{12}$ and $A_{21}$ can be observables. Something denied in the MKC-models, where by assuming that $A_{11}$ is an observable it follows that at least one of $A_{12}$ and $A_{21}$ is a pseudo observable. Cabello and Cunha prohibit this conclusion based on spatial separation between the particles. However, this is not an argument that is valid in a non-local theory. Indeed, in the MKC-model of the two-particle system almost none of the observables coincide with self-adjoint operators of the form $\een\otimes A$ or $A\otimes \een$ (the supposed `local' observables). Indeed, by the finite precision argument, the measurement of the observable $A_{11}$ is most likely to reveal the value of some non-local observable. So whatever experimental test is proposed in \cite{Cabello10}, it is more likely a test of locality than of contextuality.

Closely related to this discussion is a peculiar feature of the MKC-models first noted by Appleby in \cite{Appleby02}. If one considers a single particle, the observables that apply to this system do not coincide with the set of observables one ascribes to the particle if it is considered as a part of a system of multiple particles. From this Appleby concluded that the MKC-models exhibit a new form of contextuality, namely, existential contextuality; the existence of an observable depends on the context in which the system is viewed (i.e. a single particle system, or a multiple particle system). To the presented problem the advocate of the MKC-models may respond in two possible ways.

First, as the MKC-models are non-local, one may argue that from that point of view no isolated systems exist. The description provided for the single particle is merely an approximation, and the real observables are those that apply to the whole of all particles, i.e. every actual observable is extremely non-local. Secondly, one may argue that the observables ascribed to single particle systems \emph{are} the real observables. Then, to prevent empirical discrepancies, one is required to assume that simultaneous measurements on separate particles are impossible; one of the measurements occurs first and then causes an instantaneous state change for the other particle. Note that this is an actual modification of the MKC-models, and a violation of (\ref{Cabineq}) is obtained in an other way.


\subsection{Experimental tests for operational contextuality}\label{Spekkenstest}
In \cite{Spekkens05} suspicion of the possibility of experimental tests for operational contextuality was still an issue because 
\begin{quote}
``finite precision might imply that in practice no two experimental procedures are found to be operationally equivalent, in which case the assumption of noncontextuality is never applicable.'' 
\end{quote}
This is indeed a very practical objection and we should note that it doesn't apply specifically in the context of the MKC-models (being based on an other finite precision argument) but applies in general due to the technical notion of operational equivalence. Although this is a very strong idea, it is waved aside on the basis of a vaguely stated continuity criterion in \cite{Spekkens09}, where an experimental test for operational contextuality is proposed and performed. More specifically, it is shown that there is an upper bound to the success rate for a certain information theoretic task that holds for any operational theory that admits a preparation non-contextual hidden variable model, but that is violated by quantum mechanics. We will not criticize the proof on basis of the experimental difficulties, nor does an appeal to the MKC-models work in this case; surely they can perform the task as good as quantum mechanics, but they may be seen to be preparation contextual if one accepts the ambiguous way it is used in \cite{Spekkens05} and \cite{Spekkens09}. Instead, we will find a more basic objection.

The information theoretic task has the following setup. Alice is given a pair of bits $(a_1,a_2)$ uniformly at random and Bob is given a number $b\in\{1,2\}$ uniformly at random. Bob's task is to construct a number $X$ on the basis of the information available to him such that $\pee[X=a_b]$ is maximal, i.e. determining $\sup_X\pee[X=a_b]$.\footnote{In \cite{Spekkens09} one considers the more general case where Alice is given $n$ bits and Bob is given a number at random from the set $\{1,\ldots,n\}$. For our discussion, the case $n=2$ is rich enough to cover our objections.} If there's no information available to him, he might as well flip a coin for deciding $X$ since in that case $\sup_X\pee[X=a_b]=\tfrac{1}{2}$. On the other extreme end, if Alice is allowed to communicate any information to Bob, Bob can always determine $a_b$ correctly and simply takes $X=a_b$, i.e., in that case $\sup_X\pee[X=a_b]=1$. So the interesting cases are those where there is some restriction on the information Alice can send to Bob. The restriction introduced in \cite{Spekkens09} is that Alice can send any information as long as it doesn't contain any information about the parity of her two bits (i.e. the information must be parity-oblivious). That is, on basis of the information Bob receives he can make no better guess about the number $a_1+a_2 \mod 2$ than he could without the information. For example, it is allowed that Alice communicates only the value of $a_1$ to Bob, since this provides no information about the parity of $(a_1,a_2)$. In that case, if $b=1$, Bob can predict $a_b$ with certainty, but if $b=2$ he may only guess $a_b$ correctly with a probability of $\tfrac{1}{2}$. So for this setup we find $\pee[X=a_b]=\tfrac{3}{4}$. It is shown in \cite{Spekkens09} that for information that is classically encoded (i.e. in the form of a string of classical bits) this is also the best strategy. That is, given the restriction of parity-obliviousness, one has $\sup_X\pee[X=a_b]=\tfrac{3}{4}$. Moreover, it is shown that for any form of information-processing that is described by an operational theory that admits a preparation non-contextual hidden variable model this is the upper bound of success. It is then shown that in quantum mechanics and in practice one can do better, proving that quantum mechanics and nature are preparation contextual. We will now repeat the violation of this boundary by quantum mechanics as shown in \cite{Spekkens09} and then construct a classical hidden variable model\footnote{By `classical' we mean classical in the sense that the model is in accordance with Newtonian physics.} that reproduces this success rate, and show that with the use of hidden variables one can do even better. Together with the results in \cite{Spekkens09} this then shows that classical physics is preparation contextual.

For a given pair of bits, $(a_1,a_2)$, Alice prepares a system according to some procedure $P_{a_1a_2}$. And given a number $b\in\{1,2\}$ Bob performs a measurement $M_b$ on the system. Then, Bob processes the measurement outcome according to some rule to obtain his guess $X$ for the value of $a_b$. In the quantum mechanical case the preparation $P_{a_1a_2}$ used by Spekkens et. al. consists of preparing a single qubit in the state $\rho_{a_1a_2}$ where
\begin{equation}
\begin{aligned}
	\rho_{00}&=\frac{1}{2}\begin{pmatrix}1 & \tfrac{1}{2}\sqrt{2}(1-i)\\ \tfrac{1}{2}\sqrt{2}(1+i) & 1\end{pmatrix},\\
	\rho_{01}&=\frac{1}{2}\begin{pmatrix}1 & \tfrac{1}{2}\sqrt{2}(1+i)\\ \tfrac{1}{2}\sqrt{2}(1-i) & 1\end{pmatrix},\\
	\rho_{10}&=\frac{1}{2}\begin{pmatrix}1 & -\tfrac{1}{2}\sqrt{2}(1+i)\\ -\tfrac{1}{2}\sqrt{2}(1-i) & 1\end{pmatrix},\\
	\rho_{11}&=\frac{1}{2}\begin{pmatrix}1 & -\tfrac{1}{2}\sqrt{2}(1-i)\\ -\tfrac{1}{2}\sqrt{2}(1+i) & 1\end{pmatrix}.
\end{aligned}
\end{equation}
From the equality $\rho_{00}+\rho_{11}=\rho_{10}+\rho_{01}$ it follows that no measurement on the qubit can reveal any information about the parity of $(a_1,a_2)$ (c.f. \cite{Spekkens09}). 

Now Bob's strategy is the following. If Bob has $b=1$ he measures $\sigma_x$ and guesses $a_1=0$ if he finds the result 1, and guesses $a_1=1$ if he finds the value -1. It is then a straight forward calculation to show that for $b=1$ $\pee[X=a_b]=\tfrac{1}{2}+\tfrac{1}{4}\sqrt{2}$. Similarly, if $b=2$ Bob measures $\sigma_y$ and a similar process leads to the same success rate so $\pee[X=a_b]=\tfrac{1}{2}+\tfrac{1}{4}\sqrt{2}$ for every $b$, which is a higher success rate then the maximal one with only communicating a string of classical bits.
We will now construct a classical information processing procedure that reproduces these success rates, but violates the parity-obliviousness. Then we make an adjustment to recover parity-obliviousness while maintaining the same success rates. 

Suppose instead of a qubit, Alice sends two classical bits $(\lambda_1,\lambda_2)$ whose values are chosen at random according to the probability measures $\pee_{a_1a_2}$:
\begin{equation}\label{preparaties1}
\begin{tabular}{|c|c|c|c|c|}
\hline
$\lambda_1\lambda_2$ & $00$ & $01$ & $10$ & $11$ \\
\hline
$\pee_{00}$ & $\tfrac{3}{8}+\tfrac{1}{4}\sqrt{2}$ & $\tfrac{1}{8}$ & $\tfrac{1}{8}$ & $\tfrac{3}{8}-\tfrac{1}{4}\sqrt{2}$\\
\hline
$\pee_{01}$ & $\tfrac{1}{8}$ & $\tfrac{3}{8}+\tfrac{1}{4}\sqrt{2}$ & $\tfrac{3}{8}-\tfrac{1}{4}\sqrt{2}$ & $\tfrac{1}{8}$\\
\hline
$\pee_{10}$ & $\tfrac{1}{8}$ & $\tfrac{3}{8}-\tfrac{1}{4}\sqrt{2}$ & $\tfrac{3}{8}+\tfrac{1}{4}\sqrt{2}$ & $\tfrac{1}{8}$\\
\hline
$\pee_{11}$ & $\tfrac{3}{8}-\tfrac{1}{4}\sqrt{2}$ & $\tfrac{1}{8}$ & $\tfrac{1}{8}$ & $\tfrac{3}{8}+\tfrac{1}{4}\sqrt{2}$\\
\hline
\end{tabular}
\end{equation}
 In this example Bob takes for $X$ the value of $\lambda_b$ as a guess for the value of $a_b$. We then find for all $(a_1,a_2)$ and $b$ that $\pee[X=a_b]=\tfrac{1}{2}+\tfrac{1}{4}\sqrt{2}$ which is precisely the result we had in the quantum case. However, in this case the information send by Alice is no longer parity-oblivious since if Bob takes the value $\lambda_1+\lambda_2 \mod 2$ as a guess for the value of the parity, he has a success rate of
\begin{equation}
\pee[\lambda_1+\lambda_2=a_1+a_2 \mod 2]=\tfrac{3}{4}>\tfrac{1}{2}.
\end{equation}

To re-obtain parity-obliviousness, instead of just sending the information $(\lambda_1,\lambda_2)$, Alice wraps up this information in a system in a special way. Each of the values $\lambda_1$ and $\lambda_2$ are written on separate pieces of paper. These pieces are put in a box with two separate, closed compartments; the value of $\lambda_1$ stored in compartment 1 and the value of $\lambda_2$ in compartment 2. This box is then send to Bob. However, the box is booby-trapped: upon opening either compartment, the paper in the other compartment will be incinerated. So Bob is only allowed to either obtain the value $\lambda_1$ or $\lambda_2$, but not both. His guess for $a_b$ is still as good as it is in the quantum case, but he can no longer obtain any information about the parity of $(a_1,a_2)$. 

It may be clear that with the use of the box, this is no longer the best strategy for guessing the value of $a_b$. Indeed, if Alice just stores the values of $a_1$ and $a_2$ in the box in stead of $\lambda_1$ and $\lambda_2$, Bob can guess $a_b$ with certainty while maintaining parity-obliviousness. So in the classical case with use of the box we have $\sup_X\pee[X=a_b]=1$ which is higher than the upper bound for the quantum case. 
 
The incineration-box gives us an insight into the notion of preparation non-contextuality and its connection to parity-obliviousness. If one accepts the preparations in (\ref{preparaties1}) as the only possible preparations, one trivially has that the model is preparation non-contextual because none of the preparations are operationally equivalent. If one also admits other preparations --in the form of convex combinations-- preparation contextuality reappears. For example, the preparations 
\begin{equation}
\begin{tabular}{|c|c|c|c|c|}
\hline
$\lambda_1\lambda_2$ & $00$ & $01$ & $10$ & $11$ \\
\hline
$\pee_{0}$ & $\tfrac{3}{8}$ & $\tfrac{1}{8}$ & $\tfrac{1}{8}$ & $\tfrac{3}{8}$\\
\hline
$\pee_{1}$ & $\tfrac{1}{8}$ & $\tfrac{3}{8}$ & $\tfrac{3}{8}$ & $\tfrac{1}{8}$\\
\hline
\end{tabular}
\end{equation}	
are operationally equivalent, yet give different probabilities for the hidden variables $(\lambda_1,\lambda_2)$. Then if the booby-trap is removed one is also able to measure both $\lambda_1$ and $\lambda_2$ and preparation non-contextuality is again obeyed. However, the additional possible measurements also nullify the parity-obliviousness. So preparation non-contextuality and parity-obliviousness are strongly linked and both rely on the \emph{total of all admissible preparations and measurements}. In fact, this shows the dependence of the meaning of preparation non-contextuality on the used theoretical framework, for it is only within the framework that the sets of all admissible preparations and measurements are defined. 

The quantum mechanical case can be viewed quite similar: parity-obliviousness is only maintained by the assumption that direct measurements of the state of the system $\rho_{a_1a_2}$ are impossible. If one was able to determine the state in a single measurement, preparation non-contextuality would be obtained and parity-obliviousness would no longer be satisfied. However, the most important lesson to be learned is that preparation contextuality is not a typical quantum phenomenon; it can also appear in reasonable classical systems for which a restriction on the possible measurements exist. So although it may play a significant role in information theory (due to its link with parity-obliviousness), it has no significance in the discussion on the foundations of quantum mechanics. 


\subsection{Discussion and conclusion}
We have explained and shown that there can be no experimental test of non-contextuality (NC) under the assumption of Re and FM. We state these additional assumptions explicitly since without them the whole discussion would be trivial. For example, in the Copenhagen interpretation (rejecting Re and FM) NC can be taken as an axiom for the probabilities that the Born-rule assigns to possible measurement outcomes for an observable only depend on the self-adjoint operator with which the observable is associated. This raises the question of the physical relevance of the experimental tests. 
Rather then seeing the tests as an indication for one of the assumptions to be false, they may also be interpreted as proof for one of the assumptions, namely, QM. Indeed, the experimental tests reveal explicitly how the functional relationships between observables is respected by the measurement outcomes, suggesting that such observables are strongly correlated.\footnote{Of course, the experimental tests of non-locality may be taken to also show this.} Consequently, the tests also give us more insight into what part of the Hilbert space structure is essential for the description of certain phenomena.

The advocates of experimental tests of contextuality may of course still argue that our definition of non-contextuality is too artificial, and that the tests may exclude some other form of non-contextuality. Although the operational notion of non-contextuality was partly introduced to generalize definitions like NC to a less artificial level, we have seen that quite the opposite appears to be the case. The operational approach introduces new ambiguities with respect to operational equivalence and notions of non-contextuality that are so strong that even classical systems are deemed to be non-classical.

 We still agree that NC is an artificial criterion, but we think that any notion of non-contextuality that is more satisfying is even less likely to be the subject of empirical investigations. The definition of NC depends strongly on the formulation of orthodox quantum mechanics and thus doesn't make much sense in arbitrary theories. One may therefore propose an other definition:
\begin{itemize}
\item[\textbf{NC'}] The outcome of a measurement is independent of which other measurements are performed simultaneously.
\end{itemize}
It is clean of any reference to quantum mechanics and also doesn't include a clause on the compatibility of measurements, as is sometimes seen. In orthodox quantum mechanics and the hedgehog interpretation this criterion is satisfied since the Born rule assigns probabilities independent of what measurements are being performed (c.f. \cite[$\S$5.2]{Hermens10}, \cite[$\S$VII]{Mermin93}).

The discussion becomes more interesting if one also assumes Re and FM, since one then infers that the value of a measured observable is independent of which other observables are being measured. This is still true in the hedgehog interpretation, although one does have that the actual observable being measured may depend on which other observables are being measured. For example, in terms of the Mermin-Peres square, if one wishes to measure the (pseudo) observable $A_{11}$ simultaneously with $A_{12}$, the actual observable being measured differs from the one measured if one measures it simultaneously with $A_{21}$. One may interpret this as yet an other form of contextuality. Our response to this view reveals one of the main problems with contextuality. This form of `contextuality', that the MKC-models possess, is quite reasonable since one cannot expect in general that in different experimental setups one is able to measure the same observables. This provides an argument for any theory to escape the label of being contextual; if one observable has different values in several measuring contexts, apparently it isn't the same observable in these different contexts, and one simply introduces extra observables to obtain non-contextuality. So any theory can be made non-contextual (in the sense of NC') by introducing sufficiently many new observables.

All in all, it seems that non-contextuality can only be defined in a satisfactory way within the framework of a theory (like our definition NC for quantum mechanics), reducing the metaphysical relevance of any statement concerning such a notion of non-contextuality. 
We therefore claim that although contextuality (if well specified) can be used to compare theories for which it is well-defined, any theory-independent definition of non-contextuality is either ambiguous, or trivially satisfiable by any theory. It therefore seems a concept that is unlikely to play an unambiguous role on the metaphysical level of the study of nature, which seems the attempt of some of the authors of the experimental tests of contextuality. Or, to put this conclusion more bluntly: there is no experimental metaphysics with respect to contextuality.

On a positive note, we do believe that experimental metaphysics with respect to non-locality may well be possible. The reason for this is that whereas the notion of non-contextuality is quite counterfactual in nature (leading to the paradox we discussed in section \ref{paradox}), this isn't the case for non-locality. Of course it remains to be proven that the finite-precision argument doesn't cause a problem for the derivation of Bell-inequalities and other results that attempt to show the necessity of non-locality. We aim to provide these proofs in a future article. 

\section*{Acknowledgments}
The author would like to thank the following people for support and inspiration: A. Kent and N. P. Landsman, but most of all M. P. Seevinck for extensive commentary, useful references and discussions.

\bibliographystyle{hplain}
\bibliography{referenties}


\end{document}